\documentclass[aps,prl,onecolumn,showpacs]{revtex4-1}
\usepackage[dvips,pdftex]{graphicx}
\usepackage{amsmath}
\usepackage{color}

\begin{document}
\title{Anderson attractors in active arrays}

\author{T.V.~Laptyeva$^1$, A.A.~Tikhomirov$^2$, O.I.~Kanakov $^2$ and M.V.~Ivanchenko$^3$}

\affiliation{$^1$ Theory of Control Department, Lobachevsky State University of Nizhny Novgorod, Russia \\
$^2$ Theory of Oscillations Department, Lobachevsky State University of Nizhny Novgorod, Russia \\
$^3$ Department of Applied Mathematics, Lobachevsky State University of Nizhny Novgorod, Russia}

\begin{abstract}
In dissipationless linear media, spatial disorder induces Anderson localization of matter, light, and sound waves. The addition of nonlinearity causes interaction between the eigenmodes, which results in a slow wave diffusion. We go beyond the dissipationless limit of Anderson arrays and consider nonlinear disordered systems that are subjected to the dissipative losses and energy pumping.  We show that the Anderson modes of the disordered Ginsburg-Landau lattice possess specific excitation thresholds with respect to the pumping strength. When pumping is increased above the threshold for the band-edge modes, the lattice dynamics yields an attractor in the form of a stable multi-peak pattern. 
The Anderson attractor is the result of a joint action by the pumping-induced mode excitation,
nonlinearity-induced mode interactions, and dissipative stabilization. The regimes of Anderson attractors can be potentially realized with polariton condensates lattices, active waveguide or cavity-QED arrays.
\end{abstract}

\noindent 

\maketitle

\section*{Introduction}

%
%%%%%%% introduction
After more than fifty years since its birth, 
Anderson localization still remains in the focus of studies \cite{Evers2008,fifty}. 
During the last decade it became almost ubiquitous in experimental physics, being 
observed with electromagnetic \cite{Optics}, acoustic \cite{Hu08}, and matter waves \cite{Billy08,Roati08,Kondov11,Jendr2011}. 
In the theoretical domain, a generalized problem of localization in presence 
of nonlinearity and interactions was brought to the forefront of the studies \cite{Shepelyansky_1993,Molina1998,Pikovsky2008,Fishman2008,Flach2009,Laptyeva2010,Johansson2010,Basko_2011,Michaely2012,csigsf13}. The predicted wave packet delocalization and chaotic subdiffusion has already received an impressive support in the pioneering experiments with interacting ultracold atoms expanding in effectively one dimensional (1D) optical potentials \cite{nonlinear2,Deissler2010,Lucioni2011}.

Most of the current activity in the field remains restricted to a dissipationless limit, when the dynamics of a system is fully specified by its Hamiltonian. Otherwise, since Anderson localization is a phenomenon relying on interference \cite{And58}, one expects the destructive effect of dissipation due to rising of decoherence effects. Indeed, absorption of light in waveguide arrays (and, optionally, gain) and disorder have proved to produce an intricate interplay instead of pure Anderson localization, though permitting strongly suppressed diffusion \cite{Frank2006,Yamilov2014}. Likewise, it has been demonstrated for quantum particles that scattering \cite{Fyodorov} and spectral \cite{Huse} properties of localizing systems are deteriorated, though survive weak dissipation or coupling to a Hamiltonian bath, respectively. Noteworthy, dissipation in ordered lattices have proved to be destructive for the originally ballistic transport. Namely, it evokes the mobility transition towards diffusive light propagation, when introduced homogeneously \cite{Eichelkraut_2013}, and exponential localization, when randomized \cite{Basiri_2014}. Instructively, the dissipation introduced at the boundaries of passive chains  (or mimicked by semi-infinite propagating leads) organizes non-trivial transitions in the scaling of relaxation \cite{Kottos_2004}, transparency  \cite{Tietsche_2008}, and arising asymmetry of wave propagation \cite{Lepri_2011}, depending on the levels of disorder and nonlinearity.  

The first example of the constructive interplay was recently found in a random laser operating in the Anderson regime, when localization reduced the spatial overlap between lasing modes, preventing their competition and improving stability \cite{LiuJ.2014}. 
%There quantum wells were embedded as a gain medium in the disordered photonic crystal membrane and were optically pumped. Anderson localization reduced the spatial overlap between lasing modes, thus preventing mode competition and improving stability. 
Importantly, distinct lasing thresholds for Anderson modes in pumping strength were observed, enabling sequential excitation and control. 
It was also argued that interactions between the modes get suppressed in the strong localization and vanishing dissipation limit, although with significant deviations found beyond \cite{Stano2013}.

A new room for dissipation effects was created by the recent progress
in experimental manipulations with exciton-polariton  condensates 
\cite{Kasprzak,Balili,Deng2010,Carusotto2013,Byrnes2014}. A condensate 
can be considered as an active system balancing between excitation 
(by a pumping source) and decay (due to the continuous light emission). 
Further on, one can arrange 1D arrays of condensate centers 
by synthesizing spatial inhomogeneities  \cite{Balili, Lai, Tanese, Bloch} or by rotating 
ring-shaped optical potentials and switching to the co-moving frame \cite{amico,berloff}.
Spatial interaction appears due to polariton diffraction and diffusion and, importantly, would include both Josephson and dissipative terms (the former typically prevails). 
The resulting collective dynamics is a blend of excitation and lasing effects and can be 
modeled with Ginzburg-Landau type equations (GLE) \cite{GLE}.
In this framework, dissipative effects act as internal decay mechanisms and their 
influence on the center dynamics is accounted by additional imaginary terms 
in the model equations \cite{Deng2010,Byrnes2014,GPE,Cristofolini2013}. 
The recent pioneering theoretical and experimental studies have already demonstrated a rich nonlinear dynamics of traveling and immobile gap solitons in periodic 1D condensate center arrays \cite{Tanese,Kivshar}, and further stretched to spatially quasiperiodic structures to uncover the fractal energy spectrum \cite{Bloch}.   

Altogether, these advances naturally lead to 
the question of Anderson localization in \textit{active} arrays, where 
pumping and dissipation join the old players, nonlinearity and disorder. 
Some collective phenomena in such systems are well studied,  for example, 
synchronization \cite{Pik_book} and oscillation death \cite{Osipov1998,Rubchinsky2000,Rubchinsky2002}. However, most of the related studies 
address  lattices that crumble into a set of uncoupled oscillators 
in the linear conservative limit. 
%The nonlinear dynamics of a 
%paradigmatic class of active disordered systems with finite localization length has thus been left aside.  

In this Report we demonstrate and study Anderson attractors in 1D active arrays, as described by  
a disordered \cite{And58} version of the discrete complex GLE \cite{dGLE}. We find that the increase of the pumping strength leads to the formation of a stationary multipeak pattern
formed  by a set of excited and interacting Anderson modes. 
We determine the transition from the regime of Anderson attractors to delocalized collective oscillations
upon the increase of pumping. Both excitation and delocalization thresholds scale 
with the strength of the dissipative coupling and increase with the increase of disorder. 
Finally, we show that the increase of pumping beyond the delocalization threshold leads to a multi-mode chaos
followed by cluster synchronization.

\section*{Results}

We consider a one-dimensional disordered discrete Ginsburg-Landau equation, a generalization of the original Anderson lattice equations \cite{And58} that suitably accounts for non-equilibrium condensate dynamics \cite{Cristofolini2013}
%%%%%%%%%%%%%%%%%%%%%%%%%%%%% equation %%%%%%%%%%%%%%%%%%%%%%%%%%%%%%%%%
\begin{equation}
\label{eq:1a}
\begin{aligned}
&i\dot{z}_l=\Delta_l z_l + i\left(\alpha-\sigma\left|z_l\right|^2\right)z_l+\left|z_l\right|^2z_l-\left(1-i\eta\right)(z_{l+1}-2z_l+z_{l-1}),
\end{aligned}
\end{equation}
%%%%%%%%%%%%%%%%%%%%%%%%%%%%%%%%%%%%%%%%%%%%%%%%%%%%%%%%%%%%%%%%%%%%%%%
where $\Delta_l\in\left[-W/2, W/2\right]$  
are independent uniformly distributed random numbers and $W$ is the disorder strength.
Further on, $\alpha$ is the pumping rate, $\sigma$ is the nonlinear dissipation coefficient, 
and $\eta$ is the strength of dissipative coupling between adjacent sites. 
Without loss of generality we set conservative nonlinearity and coupling coefficients to one. In numerics, we study finite systems, and do not find appreciable finite size-effects for reasonably large array lengths, $N>100$. Zero boundary conditions are assumed for definiteness, $z_0=z_{N+1}=0$. 

In the linear dissipationless limit, $\alpha=\eta=0$ and $\left|z_l\right|^2 \rightarrow 0$,
the stationary solutions $z_l=A_l {\rm e}^{ - i \lambda t}$  satisfy
%%%%%%%%%%%%%%%%%%%%%%%%%%%%% equation %%%%%%%%%%%%%%%%%%%%%%%%%%%%%%%%%
\begin{equation}
\label{eq:1b}
\lambda_\nu A_l^{(\nu)}=\Delta_l A^{(\nu)}_l-A^{(\nu)}_{l+1}+2A^{(\nu)}_l-A^{(\nu)}_{l-1},
\end{equation}
%%%%%%%%%%%%%%%%%%%%%%%%%%%%%%%%%%%%%%%%%%%%%%%%%%%%%%%%%%%%%%%%%%%%%%%
which by $E_\nu \equiv \lambda_\nu-2$ reduces to the standard Anderson eigenvalue problem.
All eigenstates $A_l^{(\nu)}$  are exponentially localized,
$|A^{(\nu)}_l|\sim \exp\left[-|l-l_\nu|/\xi_\lambda \right]$,
%%%%%%%%%%%%%%%%%%%%%%%%%%%%% equation %%%%%%%%%%%%%%%%%%%%%%%%%%%%%%%%%
%\begin{equation}
%\label{eq:2}
%|A^{(\nu)}_l|\sim \exp\left[-\frac{|l-l_\nu|}{\xi_E}\right],
%\end{equation}
%%%%%%%%%%%%%%%%%%%%%%%%%%%%%%%%%%%%%%%%%%%%%%%%%%%%%%%%%%%%%%%%%%%%%%%
with $l_\nu$  and $\xi_\lambda$ denoting a center of mass and localization length of the mode, respectively.
The eigenvalues are restricted to a finite interval, $\lambda_\nu \in \left[-W/2, 4+W/2 \right]$. 
In the limit of weak disorder, $W \ll 1$, and far from the band edges, $0<\lambda_\nu<4$, the localization length is approximated by \cite{Thouless_1979}
%\cite{Cris93, Ish73}
%%%%%%%%%%%%%%%%%%%%%%%%%%%%% equation %%%%%%%%%%%%%%%%%%%%%%%%%%%%%%%%%
\begin{equation}
\xi_\lambda \approx  \frac{24(4-E^2(\lambda))}{W^2}=\frac{24\lambda(4-\lambda)}{W^2}.
\label{eq:1p6ppppp}
\end{equation}
%t%%%%%%%%%%%%%%%%%%%%%%%%%%%%%%%%%%%%%%%%%%%%%%%%%%%%%%%%%%%%%%%%%%%%%%%

Switching to the Anderson mode basis $z_l=\sum_\nu \psi_\nu(t) A_l^{(\nu)}$, we recast the original equation 
(\ref{eq:1a}) in the form:
%%%%%%%%%%%%%%%%%%%%%%%%%%%%% equation %%%%%%%%%%%%%%%%%%%%%%%%%%%%%%%%%
\begin{equation}
\label{eq:3}
\begin{aligned}
&i\dot{\psi}_\nu=\lambda_\nu \psi_\nu + i\left(\alpha-\eta\lambda_\nu\right)\psi_\nu+i\eta\sum_{\nu_1} J_{\nu,\nu_1} \psi_{\nu_1}+(1-i\sigma)\sum \limits_{\nu_1,\nu_2,\nu_3}I_{\nu,\nu_1,\nu_2,\nu_3}\psi_{\nu_1}\psi_{\nu_2}^\ast\psi_{\nu_3},
\end{aligned}
\end{equation}
%%%%%%%%%%%%%%%%%%%%%%%%%%%%%%%%%%%%%%%%%%%%%%%%%%%%%%%%%%%%%%%%%%%%%%%
where $J_{\nu,\nu_1}=\sum_l\Delta_l A_{l}^{(\nu)}A_{l}^{(\nu_1)}$
and $I_{\nu,\nu_1,\nu_2,\nu_3}=\sum_l A_{l}^{(\nu)}A_{l}^{(\nu_1)}A_{l}^{(\nu_2)}A_{l}^{(\nu_3)}$.
These equations contain both the linear and nonlinear terms 
that account for dissipation and pumping. Nonlinear terms are responsible for the mode interaction. 
However, due to the exponential localization of the eigenstates, 
interactions are  confined to localization volume $V_{loc}(\lambda) \approx 3.3 \xi_\lambda$ \cite{Krim10}.

We start the analysis of  Eq. (\ref{eq:1a}) by considering the net norm 
$Z=\sum |z_l|^2$. The dynamics of the  norm  is given by
%%%%%%%%%%%%%%%%%%%%%%%%%%%%% equation %%%%%%%%%%%%%%%%%%%%%%%%%%%%%%%%%
\begin{equation}
\label{eq:4}
\dot{Z}=2\sum\left[(\alpha-\sigma|z_l|^2)|z_l|^2-\eta|z_{l+1}-z_l|^2\right].
\end{equation}
%%%%%%%%%%%%%%%%%%%%%%%%%%%%%%%%%%%%%%%%%%%%%%%%%%%%%%%%%%%%%%%%%%%%%%%
It follows that the zero solution $z_l \equiv 0$ is globally stable for all $\alpha \le 0$.
It also suggests that   homogeneous in-phase solutions  $z_{l+1}\approx z_l$ 
are more energetically favorable than anti-phase ones, $z_{l+1}\approx -z_l$. 
%and, once excited, get saturated at comparatively higher amplitudes. 
To study stability of the zero solution, we  assign
increments $p_\nu$ to the small-amplitude Anderson modes, 
$z_l(t)=\zeta A_l^{(\nu)}\exp[(p_\nu-i\lambda_\nu) t], \zeta\ll1$, and substitute them into Eq. (\ref{eq:1a}). 
Linearization gives
\begin{equation}
\label{eq:5}
p_\nu=\alpha-\eta\sum\left|{A}^{(\nu)}_{l+1}-{A}_l^{(\nu)}\right|^2.
\end{equation}
The  zero solution is stable when $\mbox{max} \ p_\nu <0$.
This quantity  depends only on the  strength $W$ and particular realization $\{\Delta_l\}$ of the disorder, 
and also on the ratio between incoherent pumping rate and dissipative coupling, $\bar{\alpha}=\alpha/\eta$. 
Irrespective of the  strength and particular realization of disorder, the scaled excitation threshold 
\begin{equation}
\label{eq:6}
\bar{\alpha}^*=\min\limits_{\nu}\bar{\alpha}^*_\nu=\min\limits_{\nu}\sum\left|{A}^{(\nu)}_{l+1}-{A}_l^{(\nu)}\right|^2
\end{equation} 
is bounded, $0\le\bar{\alpha}^*\le4$. 
As the Anderson modes have finite localization lengths 
for finite disorder strength $W$ and, hence, inside localization volume we have $|A_l^{(\nu)}|\sim 1/\sqrt{V_{loc}}$, there is a finite excitation threshold $\bar{\alpha}>0$ for finite $W$.

\begin{figure}[h]
\begin{center}
\includegraphics[angle=270,width=\columnwidth,keepaspectratio,clip]{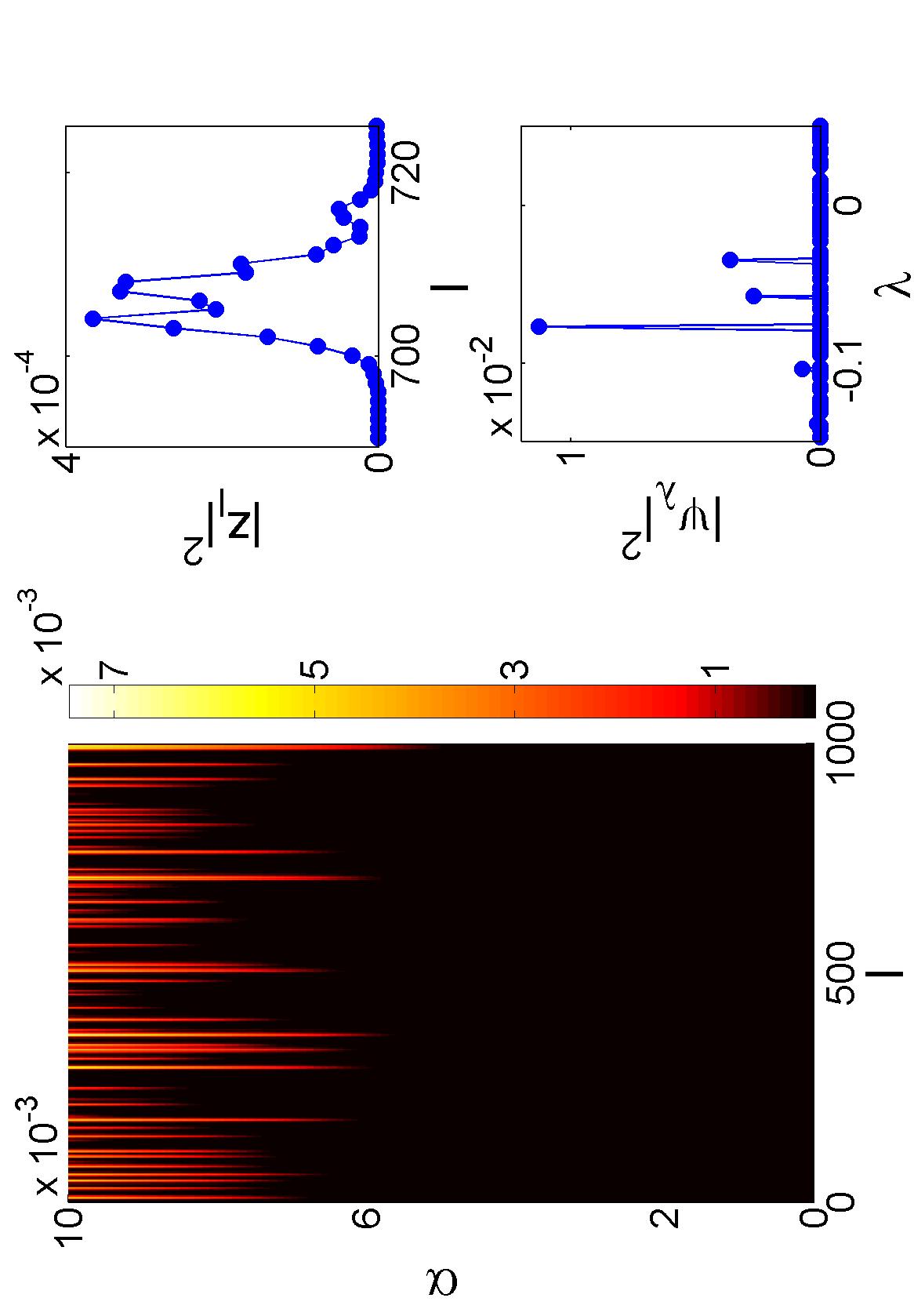}
\caption{Development of the Anderson attractor for a particular disorder realization
of the system (\ref{eq:1a}) upon the increase of the pumping. 
Left panel:  oscillation amplitudes at lattice sites, $|z_l|^2$ (color), as functions of $\alpha$. 
Profile of a single excitation spot in the direct (top right) 
and  Anderson mode space (bottom right) for $\alpha=0.006$. 
The parameters are  $W=1, \eta=0.1, \sigma=1$, $N=1000$.}
\label{fig:1}
\end{center}
\end{figure}

Figure~\ref{fig:1} presents the results of numerical simulations for a particular  realization of disorder. 
Profiles for different values of $\alpha$  were obtained as independent \textit{attractor} solutions,
by setting the system into an initial random low-energy state $|z_l(0)| \ll 1$ and letting it evolve
until the corresponding  amplitude profile is stabilized. (We observed single-attractor regimes in all performed numerical tests, although multistability is not excluded, in principle.) The key feature of the attractor patterns 
is the multipeak structure, well pronounced above a certain threshold (e.g. $\alpha \approx 0.006$ for $W=1, \eta=0.1$, $\sigma=1$, $N=1000$,  Fig.~\ref{fig:1}). The positions of  the peaks remain unaffected by the 
further increase of the pumping strength.
Zooming into a single peak, we find that it extends over many sites, top right panel of Fig.~\ref{fig:1}. 
By going into the reciprocal Anderson space, we find that the excitation
is well-localized at a single Anderson mode, bottom right panel of Fig.~\ref{fig:1}.
This observation supports the conjecture that the attractor peaks are produced
through excitation of Anderson modes.

Mode specific excitation conditions can be further analyzed by using the linearized version of equations (\ref{eq:3}),  
\begin{equation}
\label{eq:7}
i\dot{\psi}_\nu=\lambda_\nu \psi_\nu + i\left(\alpha-\eta\lambda_\nu\right)\psi_\nu+i\eta\sum_{\nu_1} J_{\nu,\nu_1} \psi_{\nu_1}.
\end{equation}

In the weak disorder limit, $W\ll1$, the
localization length of the modes that are far from the band edges is large,  $\xi_\lambda\gg1$.
Since within the localization volume $|A_l^{(\nu)}|\ll1$, the terms with $J_{\nu,\nu_1}$ can be neglected.
It follows immediately that the rescaled excitation threshold of $\nu$-th Anderson mode can be approximated well by its eigenvalue,   
%%%%%%%%%%%%%%%%%%%%%%%%%%%%% equation %%%%%%%%%%%%%%%%%%%%%%%%%%%%%%%%%
\begin{equation}
\label{eq:8}
\bar{\alpha}^*_\nu\approx\lambda_\nu.
\end{equation}
%%%%%%%%%%%%%%%%%%%%%%%%%%%%%%%%%%%%%%%%%%%%%%%%%%%%%%%%%%%%%%%%%%%%%%%
This also  means that the modes closer the lower band edge will be excited first. 
However, the localization length of such modes can substantially decrease, potentially, up to $\xi_\lambda \sim 1$, so that corrections to Eq.~(\ref{eq:7}) due to $J_{\nu,\nu_1}$ terms 
might become significant. 

The instability threshold can be estimated more accurately by using Eq.~(\ref{eq:6}). 
Neglecting exponentially decaying tails of the modes, $A_l=0, \ l\notin[l_\nu-V_{loc}/2,l_\nu+V_{loc}/2]=0$, and minimizing $\bar{\alpha}^*_\nu$ under normalization constraint $\sum A_l^2=1$, we obtain:
\begin{equation}
\label{eq:5b}
\min\bar{\alpha}^*_\nu=4\sin^2\frac{\pi}{2(V_{loc}+1)}.
\end{equation}
Finally, by substituting the localization length $\xi_0\approx 8 W^{-2/3}$ 
for the modes with $\lambda_\nu\approx0$ \cite{Derrida} in $V_{loc}\approx3.3\xi_\lambda$, we arrive at:
\begin{equation}
\label{eq:5c}
\bar{\alpha}^*\approx W^{4/3}/64.
\end{equation} 
Note, that this approach is also valid in the strong disorder limit, $W\gg1$, when all Anderson modes are essentially single-site excitations: substituting $V_{loc}=1$ in (\ref{eq:5b}) one obtains $\bar{\alpha}^*_\nu\approx2$. Moreover, taking into account the strong decay of the mode amplitudes, $\left|A_{l_\nu\pm (l'+1)}^{(\nu)}/A_{l_\nu\pm l'}^{(\nu)}\right|\sim \exp\left(-\xi_{\lambda_\nu}^{-1}\right)\ll 1$, one finds that the mode specific excitation thresholds (\ref{eq:6}) are approximated by
\begin{equation}
\label{eq:5d}
\bar{\alpha}^*_\nu\approx2+\sum\limits_{\nu\neq l_\nu}\left(A_l^{(\nu)}\right)^2\approx2(1+e^{-2/\xi_\nu}).
\end{equation}
It follows, that they tend to the limiting value $\bar{\alpha}^*=2$ as $W\rightarrow\infty$.

\begin{figure}[t]
\begin{center}
\includegraphics[angle=270,width=\columnwidth,keepaspectratio,clip]{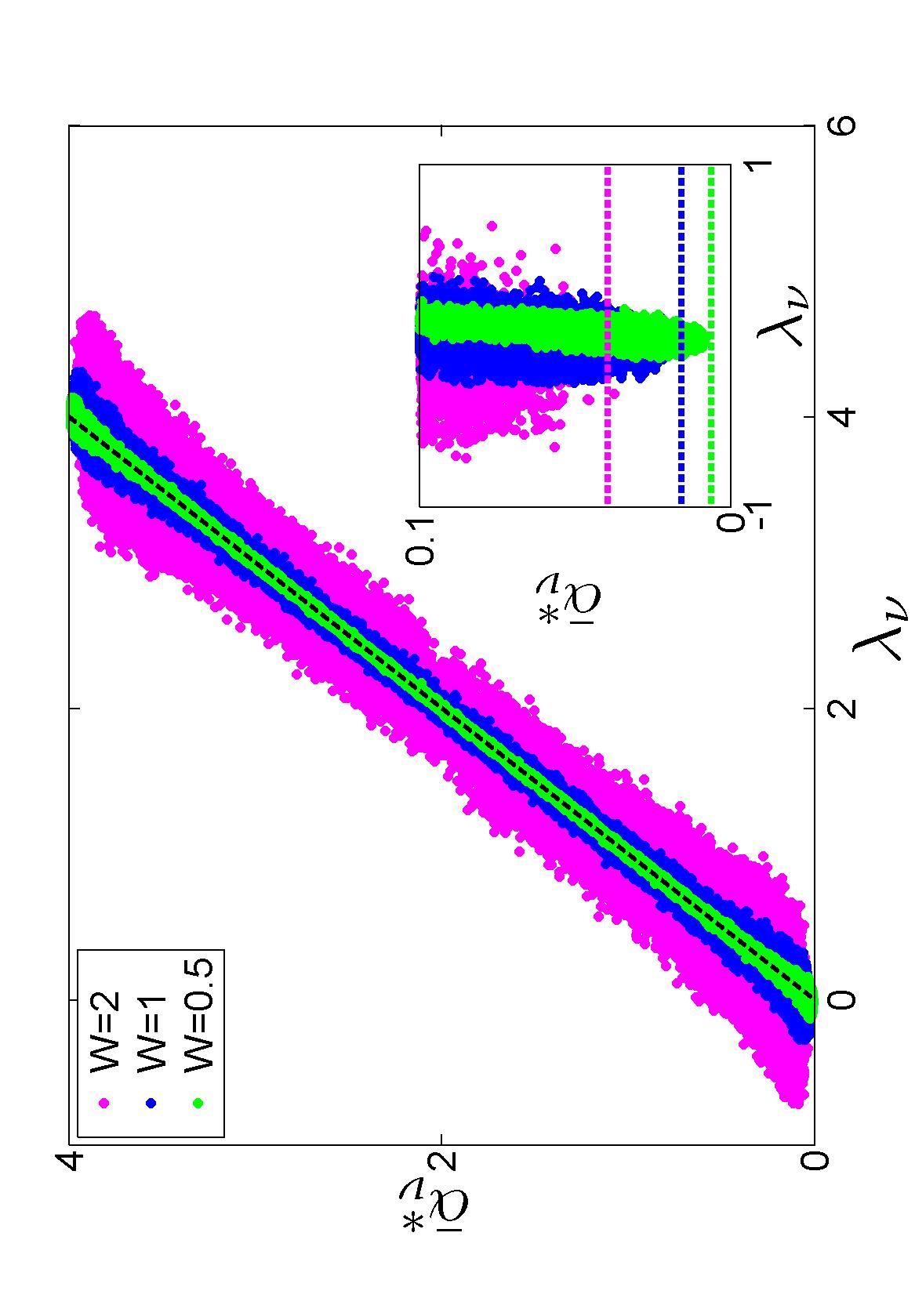}
\caption{ Rescaled mode excitation thresholds $\bar{\alpha}^*_\nu$, Eq. (\ref{eq:6}), vs
mode eigenvalues $\lambda_\nu$. Eigenvalues were obtained by numerically solving eigenvalue problem (\ref{eq:1b})
for the lattices of the size $N=1000$ and particular realizations of disorder of the strength $W=0.5$ (green), $1$ (blue), $2$ (magenta). 
Dashed line corresponds to $\bar{\alpha}^*_\nu=\lambda_\nu$. 
Inset: Zoomed fragment of the main plot. Dashed lines 
indicate   excitation thresholds obtained from Eq. (\ref{eq:5c}).}
\label{fig:2}
\end{center}
\end{figure}

To test the analytical results, we calculate  mode excitation thresholds $\bar{\alpha}^*_\nu$ according to (\ref{eq:6}) and plot them as a function of the numerically calculated eigenvalues $\lambda_\nu$, Fig.~\ref{fig:2}. The obtained statistical dependencies corroborate  approximation (\ref{eq:8}) 
for the modes far from the band edges, especially well in the limit of weak disorder. 
The values of minimal excitation thresholds correspond to $\lambda_\nu\approx0$, and the estimate (\ref{eq:5c}) 
is in a good agreement with numerical results, see inset of Fig.\ref{fig:2}. 
By approximating the dependence around its dip by $|\bar{\alpha}-\bar{\alpha}^*|\propto|\Delta\lambda|^2$ 
and taking into account the finiteness of the density of Anderson states 
at $\lambda=0$, we get that the density of excited states scales  $\propto \sqrt{\bar{\alpha}-\bar{\alpha}^*}$.

By getting over the oscillation threshold $\bar{\alpha}^*$  one would not
immediately excite all modes near the band edge. 
These modes are well-localized and their interaction with  other modes is exponentially weak. 
In addition, next-neighbor mode interaction  remains significantly damped since 
mode eigenvalues  differ substantially due to the level repulsion. 

As a  result,  Anderson modes from the vicinity of the band edge 
arise in a one-by-one manner as the pumping rate exceeds  thresholds $\bar{\alpha}>\bar{\alpha}_\nu^*$. 
Mode amplitudes  saturate because of the nonlinear dissipation and amplitude asymptotic values
can be estimated, by using Eq.~(\ref{eq:3}), as:
%%%%%%%%%%%%%%%%%%%%%%%%%%%%% equation %%%%%%%%%%%%%%%%%%%%%%%%%%%%%%%%%
\begin{equation}
\label{eq:6a}
|\psi_\nu|\approx\sqrt\frac{\alpha+\eta\lambda_\nu-\eta J_{\nu,\nu}}{\sigma I_{\nu,\nu,\nu,\nu}}.
\end{equation}
%%%%%%%%%%%%%%%%%%%%%%%%%%%%%%%%%%%%%%%%%%%%%%%%%%%%%%%%%%%%%%%%%%%%%%%
%It is worth mentioning that the effects of mode interactions on their excitation and values of saturated amplitudes were noticed in numerics, but we leave it for a separate study. 

\begin{figure}[t]
\begin{center}
\includegraphics[angle=270,width=\columnwidth,keepaspectratio,clip]{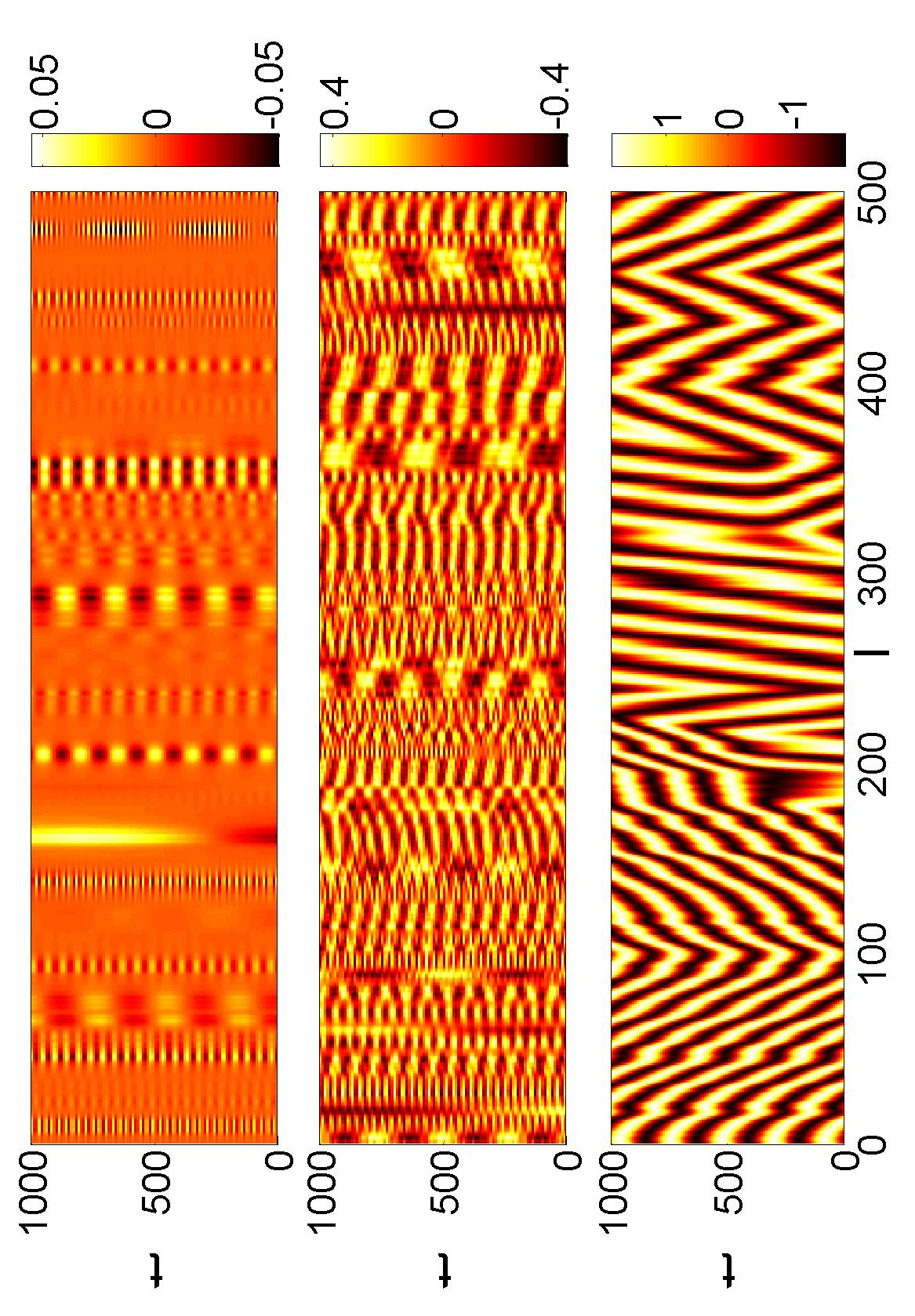}
\caption{Spatio-temporal patterns of $\mbox{Re}(z_l)$ (color) for different pumping rates:
 $\alpha=0.0075$ (top), $\alpha=0.1$ (middle), and $\alpha=3.0$ (bottom).
The profiles illustrate three different regimes:  Anderson attractor (top), 
mode chaos (middle), and cluster synchronization (bottom). The parameters are $W=1, \eta=0.1, \sigma=1$, $N=500$.}
\label{fig:4}
\end{center}
\end{figure}

%===================================== SECTION =================================================================
As the pumping strength increases further, the set of excited modes becomes dense and mode interaction starts contributing to the formation of the system attractor.  Multi-mode nonlinear dynamics 
has two well-known trademarks: chaos and synchronization \cite{Pik_book}. 
Both appear in our model system, see Fig.~\ref{fig:4}. 
By gradually increasing the pumping strength, we first observe a transition from the Anderson attractors 
to the regime of delocalized oscillations, Fig.~\ref{fig:4} (middle panel).  
The delocalized regime is characterized by irregular spatio-temporal patterns. In terms of the localized modes, this is a well-developed 
mode chaos. When  the pumping is increased further, we observe formation of synchronization clusters 
with the typical size of the Anderson localization length. 

We can estimate the transition to delocalized oscillations by assuming that it happens when the sum of the localization volumes of the excited modes becomes of the order of the system size, $\sum V_{loc}\sim\mathcal{O}(N)$. An average localization volume that measures the ratio of effectively 
excited sites is $\langle V_{loc}\rangle\sim\mathcal{O}(1)$, where the non-excited modes are formally assigned $V_{loc}=0$. By using expression (\ref{eq:8}) for the mode excitation thresholds, neglecting contributions 
of the highly localized modes near the lower band edge, and approximating  the density of states 
in the weak disorder limit as $\rho(\lambda)\approx (\pi\sqrt{\lambda(4-\lambda)})^{-1}$, we obtain
%%%%%%%%%%%%%%%%%%%%%%%%%%%%% equation %%%%%%%%%%%%%%%%%%%%%%%%%%%%%%%%%
\begin{equation}
\label{eq:13}
\langle V_{loc} \rangle\approx\int\limits^{\bar{\alpha}}_0 V_{loc}(\lambda)\rho(\lambda)d\lambda\approx\frac{105\bar{\alpha}^{3/2}}{\pi W^2}
\end{equation}
%%%%%%%%%%%%%%%%%%%%%%%%%%%%%%%%%%%%%%%%%%%%%%%%%%%%%%%%%%%%%%%%%%%%%%%
and get the transition value:
%%%%%%%%%%%%%%%%%%%%%%%%%%%%% equation %%%%%%%%%%%%%%%%%%%%%%%%%%%%%%%%%
\begin{equation}
\label{eq:14}
\bar{\alpha}^{**}\approx \left(\frac{\pi}{105}\right)^{2/3}W^{4/3}.
\end{equation}
%%%%%%%%%%%%%%%%%%%%%%%%%%%%%%%%%%%%%%%%%%%%%%%%%%%%%%%%%%%%%%%%%%%%%%%
%Here we also assumed $\bar{\alpha}^{**}\ll 6$, which always holds as we see, comparing it to the upper bound (\ref{eq:11a}).
%In particular, $ \bar{\alpha}^{**}\approx 0.31 \mbox{ for } W=2$ and $\bar{\alpha}^{**}\approx 0.125 \mbox{ for } W=1$.
In the strong disorder limit the mode excitation thresholds (\ref{eq:5d}) converge to $\bar{\alpha}^*=2$, which, therefore, also approximates the onset of delocalized oscillations, $\bar{\alpha}^{**}\approx2$. 

\begin{figure}[t]
\begin{center}
\includegraphics[angle=270,width=\columnwidth,keepaspectratio,clip]{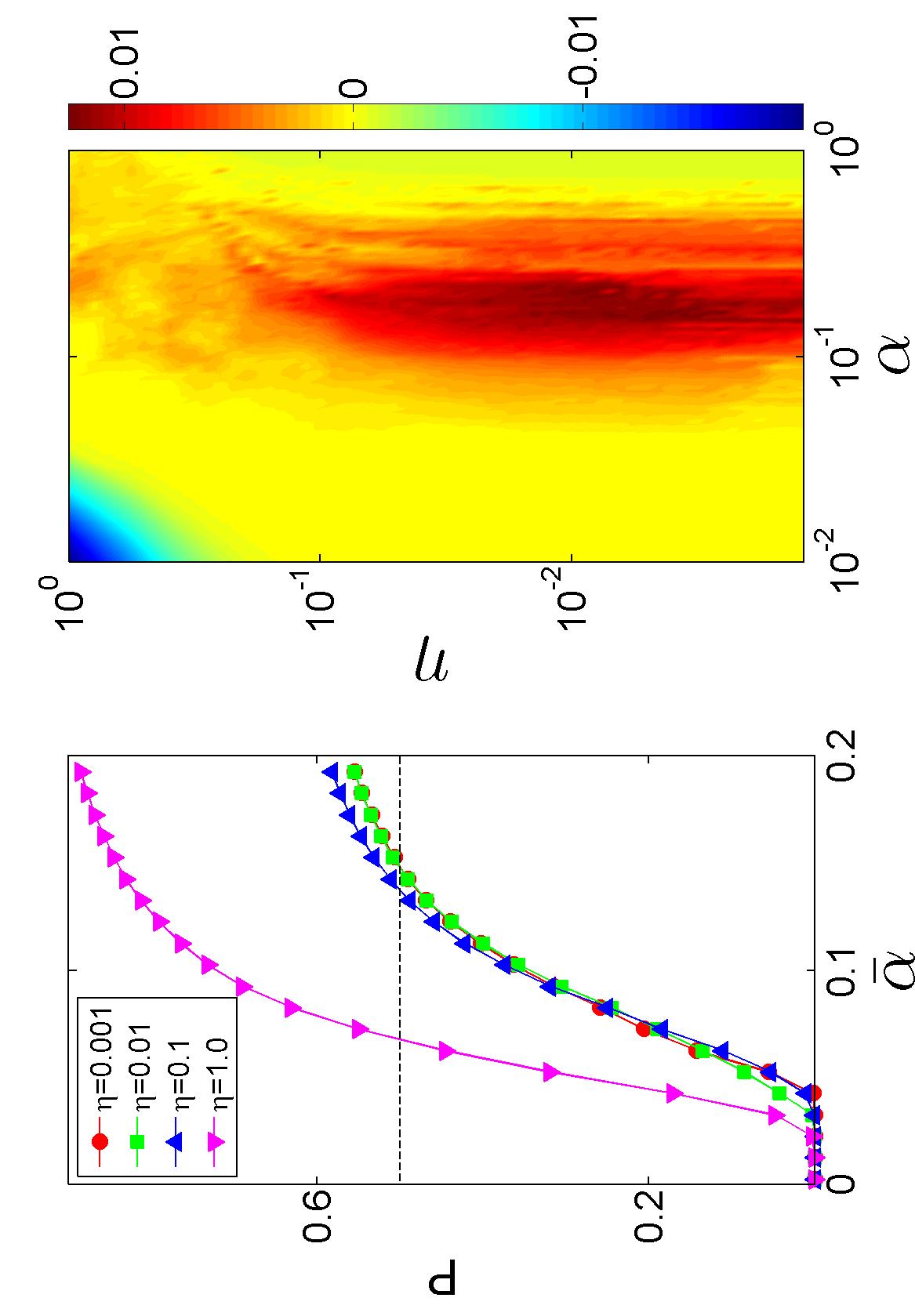}
\caption{Left panel: 
Normalized participation number, Eq.~(\ref{eq:16}), 
for the attractor of the system (\ref{eq:1a}) vs 
scaled pumping rate $\bar{\alpha}$ for different dissipative coupling strengths. 
Dashed line corresponds to $P=0.5$. 
Right panel: Largest Lyapunov exponent of the attractor (color) as a function of
the pumping $\alpha$ and dissipative coupling constant $\eta$.
The parameters are  $W=1$, $\sigma=1, N=200$. Note the difference between the  scaled $\bar{\alpha}=\alpha/\eta$ (left panel)
and non-scaled $\alpha$ (right panel) pumping constants.}
\label{fig:3}
\end{center}
\end{figure}

For a numerical test we average $|z_l|^2$ over observation time and calculate the participation number (a quantity commonly used to estimate the number of effectively excited sites) normalized by the system size:
%%%%%%%%%%%%%%%%%%%%%%%%%%%%% equation %%%%%%%%%%%%%%%%%%%%%%%%%%%%%%%%%
\begin{equation}
\label{eq:16}
P=\left(\frac{1}{N}\sum |z_l|^4/Z^2\right)^{-1}.
\end{equation}
%%%%%%%%%%%%%%%%%%%%%%%%%%%%%%%%%%%%%%%%%%%%%%%%%%%%%%%%%%%%%%%%%%%%%%%
Since the maximally possible $P=1$ requires a 
uniform distribution of $|z_l|$, we use $P=1/2$ as the threshold value to indicate localization-delocalization transition. 
The left panel of Fig.~\ref{fig:3} presents the results obtained by averaging over ten disorder realizations.
%ing dependenaces on the scaled pumping $\bar\alpha$ for disorder strength $W=1$ and the system size $N=200$. 
For weak dissipative coupling $\eta\ll1$, the scaled curves $P(\bar{\alpha})$ fall closely to each other, in accord to the theoretical prediction, Eq.~(\ref{eq:14}). It also estimates the numerical thresholds reasonably well, e.g. compare $\bar{\alpha}^{**}\approx0.1$ for $W=1$, Eq.~(\ref{eq:14}), to $\bar{\alpha}^{**}\approx0.13\ldots0.15$, as read from Fig.~\ref{fig:3}. 
When the dissipative coupling becomes of the order of the conservative one, $\eta=\mathcal{O}(1)$, 
estimate (\ref{eq:14}) with the scaling $P(\alpha,\eta,W)=P(\bar\alpha,W)$ are no longer valid, and the actual delocalization threshold is significantly different from (\ref{eq:14}). 
In this limit one cannot  neglect the last term in Eq.~(\ref{eq:7}) 
which is responsible for dissipative interaction between the  modes. 

In order to quantify the transition to the mode chaos regime, we calculate the largest Lyapunov exponent 
as a function of the pumping strength, Fig. \ref{fig:3} (right panel).
Comparing the exponents, obtained for different values of dissipative coupling constant $\eta$, with
the results presented in  Fig. \ref{fig:3}, we confirm that the transition to delocalized oscillations 
is a precursor of the mode chaos. Remarkably, a further increase of the pumping above $\alpha\approx1$ 
leads to the drop of the largest Lyapunov exponents to zero thus marking the transition back to 
regular dynamics. This transition is weakly dependent of $\eta$  and corresponds to
 the emergence of synchronized clusters \cite{Pik_book}, see  Fig. \ref{fig:4} (bottom panel).

\section*{Discussion}

Anderson localization in active disordered systems is a combined effect
produced by the energy pumping, dissipation and nonlinearity. It results in the formation of the 
Anderson attractor consisting of many localized weakly-interacting modes. 
We have found that the pumping excitation thresholds for the Anderson modes are mode-specific
and those with lowest values correspond to the modes located near the lower band edge. 
Sequential excitation of Anderson modes by tuned pumping leads to the transition from Anderson
attractors to the mode chaos and attractor patterns in the form of delocalized oscillations.   

These results pose a broad range of theoretical challenges, as studying Anderson attractors in higher dimensions, which allow for a mobility edge or criticality, in other types of localizing potentials, and their counterparts in open quantum systems. It would also be of interest to consider non-uniform dissipation, e.g. absorbing boundaries only. For the experimental perspective, lattices of exciton-polariton condensates %\cite{Balili,Lai,Bloch,Tanese,Kivshar} 
and active waveguide arrays %\cite{LiuJ.2014} 
are most promising candidates for the realization of Anderson attractors. The recent study of another localizing -- quasiperiodically modulated -- $1$D polariton condensate arrays has paved a way \cite{Bloch}, and the on-chip random lasing in the Anderson regime is, probably, the first already present example \cite{LiuJ.2014}.   Other candidates (although on the model level at the moment) are  cavity-QED arrays with the cavities  filled up with two-level atoms or qubits, where the dynamics the mean-field states in the adjoint cavities can be described by using GLE-type equation \cite{Sedov2012,Chen2012} and plasmonic nanostructures \cite{Shi2014}.   
Finally, Anderson attractor regimes can be generalized to the systems of coupled disordered Josephson 
junction arrays, marked by the recent rise of interest to dissipative response effects \cite{Basko}.

\bibliography{literatura}

\section*{Acknowledgments}

T.L. and M.I. acknowledge support of Ministry of Education and Science of the Russian Federation (Research Assignment No. 1.115.2014/K). A.T. and O.K. acknowledge support of Ministry of Education and Science of the Russian Federation (Agreement No. 02.B.49.21.0003). T.L. also acknowledges support of Dynasty Foundation.

\section*{Author contributions statement}

T.L. and M.I. conceived the study, developed the theory, analyzed the results, and wrote the manuscript.  A.T. and O.K. conducted the numerical experiments, and analyzed the results.  All authors reviewed the manuscript. 

\section*{Additional information}

The Authors declare no competing financial interests.

%%%%%%%%%%%%%%%%%%%%%%%%%%%%%%% figure %%%%%%%%%%%%%%%%%%%%%%%%%%%%%%%
%%%%%%%%%%%%%%%%%%%%%%%%%%%%%%%%%%%%%%%%%%%%%%%%%%%%%%%%%%%%%%%%%%%%%

%%%%%%%%%%%%%%%%%%%%%%%%%%%%%%% figure %%%%%%%%%%%%%%%%%%%%%%%%%%%%%%%
%%%%%%%%%%%%%%%%%%%%%%%%%%%%%%%%%%%%%%%%%%%%%%%%%%%%%%%%%%%%%%%%%%%%%

%%%%%%%%%%%%%%%%%%%%%%%%%%%%%%% figure %%%%%%%%%%%%%%%%%%%%%%%%%%%%%%%
%%%%%%%%%%%%%%%%%%%%%%%%%%%%%%%%%%%%%%%%%%%%%%%%%%%%%%%%%%%%%%%%%%%%%

%%%%%%%%%%%%%%%%%%%%%%%%%%%%%%% figure %%%%%%%%%%%%%%%%%%%%%%%%%%%%%%%
%%%%%%%%%%%%%%%%%%%%%%%%%%%%%%%%%%%%%%%%%%%%%%%%%%%%%%%%%%%%%%%%%%%%%

\end{document}